\documentclass[12pt,preprint]{aastex}

\slugcomment{AJ, in press}
\shorttitle{Photometric Identification of Cool WDs}
\shortauthors{Kilic et al.}

\begin{document}
\title{Photometric Identification of Cool White Dwarfs\footnote{Based on observations obtained with the Hobby-Eberly Telescope, which is a joint project of the University of Texas at Austin, the Pennsylvania State University, Stanford University, Ludwig-Maximilians-Universit\"at M\"unchen, and Georg-August-Universit\"at G\"ottingen.}}

\author{M. Kilic\altaffilmark{2,3}, D. E. Winget and Ted von Hippel}
\affil{The University of Texas at Austin, Department of Astronomy, 1 University Station 
C1400, Austin, TX 78712, USA}
\email{kilic@, dew@, ted@astro.as.utexas.edu}

\and

\author{C. F. Claver\altaffilmark{4}}
\affil{Kitt Peak National Observatory, National Optical Astronomy Observatory, P.O. Box 26732, Tucson, AZ 85726, USA}
\email{cclaver@noao.edu}

%

\altaffiltext{2}{Visiting Astronomer, Kitt Peak National Observatory, National Optical Astronomy 
Observatory, which is operated by the Association of Universities for Research in Astronomy, Inc. 
(AURA) under cooperative agreement with the National Science Foundation.}
\altaffiltext{3}{Visiting Astronomer, Cerro Tololo Inter-American Observatory, National Optical Astronomy
Observatory, which is operated by the Association of Universities for Research in Astronomy, Inc.
(AURA) under cooperative agreement with the National Science Foundation.}
\altaffiltext{4}{Kitt Peak National Observatory, National Optical Astronomy Observatory, which is 
operated by the Association of Universities for Research in Astronomy, Inc. (AURA) under cooperative 
agreement with the National Science Foundation.}

\begin{abstract}
 We investigate the use of a narrow-band DDO51 filter for photometric identification of cool
white dwarfs. We report photometric observations of 30 known cool white dwarfs with temperatures
ranging from 10,000 K down to very cool temperatures ($\leq3500$ K). Follow-up 
spectroscopic observations of a sample of objects selected using this filter and our photometric
observations show that 
DDO51 filter photometry can help select cool white dwarf candidates for follow-up
multi--object spectroscopy by
rejecting 65 \% of main sequence stars with the same broad--band colors as the
cool white dwarfs. This technique is not selective enough to
efficiently feed single--object spectrographs.
We present the white dwarf cooling sequence using this filter. Our observations
show that very cool white dwarfs form a sequence in the $r-DDO$ vs. $r-z$ color--color diagram
and demonstrate that significant improvements are needed in white dwarf model atmospheres.
\end{abstract}


\section{Introduction}
White dwarf stars, remnants of the earliest and all subsequent generations of star formation, are
tracers of the age and evolution of the Galaxy. 
They are initially hot and consequently
cool rapidly, though the cooling rate slows as their temperature drops,
allowing the oldest white dwarfs to remain visible. Because the cooling rate
slows, any census finds more and more white dwarfs at lower and lower temperatures
(and luminosities) until, quite abruptly, we find no more of them.
Such a census is called the white dwarf luminosity function.
Attempts to exploit the white dwarfs as chronometers showed that the white dwarf
luminosity function was a map of the history of star formation in the disk, and that there
was a shortfall of low luminosity white dwarfs -- the inevitable consequence of the finite
age of the disk (Liebert 1979; Winget et al. 1987; Liebert, Dahn \& Monet 1988).

The cool end of
the white dwarf luminosity function was estimated from 43 objects found
in the Luyten Half Second Proper Motion Survey (Luyten 1979; Liebert, Dahn \& Monet 1988).
Proper motion surveys are the most common method of searching for white dwarfs.
Since the white dwarfs are intrinsically faint, they must be close to be seen;
therefore they tend to have higher proper motions than most other stars with similar magnitudes.
Proper motion surveys cannot detect white dwarfs with small
tangential velocities, however. Therefore they have complicated and hard to
quantify completeness problems.
Wood \& Oswalt (1998) argue that the ages inferred from the Liebert, Dahn \& Monet (1988)
white dwarf luminosity function
must be considered uncertain by 15\% from sampling statistics alone.
More importantly, depending on how the data are binned, as many as 3
or as few as 1 of the 43 objects occupy the last bin in the white dwarf luminosity function,
precisely the point where all of the age leverage resides.  A recent
sample of white dwarfs in wide binaries (Oswalt et al. 1996) shows a somewhat lower luminosity
downturn which, at the 2$\sigma$ level, is consistent with no downturn at
all in the coolest bin.  The simple fact is the fainter age-dependent
end of the white dwarf luminosity function is not yet satisfactorily constrained
by observation or theory.

An investigation of the cool end of the white dwarf luminosity function
that is focused on disentangling theoretical uncertainties in the cooling process
would greatly benefit from 
a much larger kinematically unbiased sample of cool white dwarfs.
The details of the constituent input physics can affect the implied
ages of white dwarfs below $log(L/L_{\sun})\sim-4.2$ by as much as 2--3
Gyr, hence are critical for using white dwarfs as chronometers
(Hawkins \& Hambly 1999; Montgomery et al. 1999; Salaris et al. 2000).

A magnitude-limited, kinematically-unbiased sample of white dwarfs can be obtained through
a photometric survey. A unique color signature is necessary to photometrically identify
a white dwarf among the many other field stars.
The magnitude limit of a survey is also a critical factor in the search
for cool white dwarfs; if the survey cannot provide sufficiently high signal to noise ratio
data for $M_{\rm V}\sim16$, it cannot recognize cool low luminosity white dwarfs.
Broad-band photometric surveys can be used to find hot white dwarfs due to their
blue colors.
Recently, Kleinman et al. (2004) found 2551 new white dwarfs with $T_{\rm eff}\geq8000$ K 
in the Sloan Digital Sky Survey Data Release 1.
Unfortunately, the broad-band colors of cool white dwarfs are identical to the
metal poor subdwarfs.
The lack of discovery of cool white dwarfs in the Sloan
Digital Sky Survey emphasizes the fact that cool white dwarfs are
indistinguishable from subdwarfs and main sequence stars in broad-band photometric observations (e.g. Claver 1995).

In this paper, we investigate the use of the narrow-band DDO51 filter for photometric
identification of cool white dwarfs. 
We present imaging data and follow-up spectroscopy of nine cool white dwarf
candidates in \S 2. In \S 3, our observations of 30 known cool white dwarfs
including four ultra cool white dwarfs are discussed. We also show the cooling sequence
for these white dwarfs. We discuss the efficiency and possible use of this filter for
photometric identification of cool white dwarfs in \S 4 along with the observed blue
turn-off of very cool white dwarfs.

\section{Forward Approach: Photometry to Spectroscopy}

Broad-band filter photometry has a limited capacity to distinguish metal poor subdwarfs from cool
white dwarfs.
In the absence of significant line
blanketing, both the white dwarfs and subdwarfs have broad-band colors that
closely approximate those of a blackbody.
However, by comparing the flux
through a magnesium absorption line--centered filter, e.g. DDO51, several 
authors have 
suggested that cool white dwarfs could be distinguished from other field stars of similar 
$T_{\rm eff}$ (Claver 1995; Harris et al. 2001; Kilic et al. 2003). 
This is because the majority of cool white dwarfs have essentially 
featureless spectra around 5150 \AA , where subdwarfs and main sequence stars show significant 
absorption from the Mgb triplet and/or MgH. Figure 1 shows a template spectra for a K5V star (Pickles 1998) and
a 5000 K blackbody spectrum along with 
the tracing of the DDO51, r, and z filters.
It is clear from this figure that white dwarfs should be
distinguishable from the subdwarf stars using the narrow-band DDO51 filter and a combination
of broad-band filters. We note that relative to a blackbody, the K star spectrum deviates
both above and below the blackbody line- depending upon the wavelength sampled. Thus, the color
indices could just as well be affected by features in the K star beyond 5150 \AA\ as at that
wavelength. However, Mg absorption is the strongest feature in the range sampled by the chosen filters, r, z, and
DDO51.

In order to test the above claim, Ed Olsewzki kindly provided us with DDO51 photometry of
an area of 2 square degrees from the Spaghetti Survey (Morrison et al. 2000)
which overlaps with 
the Sloan Digital Sky Survey fields. A color--color diagram for this field is shown in Figure 2. 
Two hot white dwarfs found by the Sloan Digital Sky Survey are shown as open circles. Spectroscopically 
identified QSOs and stars (which are not white dwarfs) are shown as open squares and filled triangles, respectively.
Spectral IDs and photometric data for these objects are given in Table 1.
A typical errorbar for these objects is shown in the lower left corner of the figure.
White dwarfs are expected to be separated from main sequence stars in this
color--color diagram (see Figure 4.11 of Claver 1995); we have selected stars that deviate from the main
sequence as possible cool white dwarf candidates.
Cool white dwarf 
candidates selected for follow--up spectroscopy at the 9.2m Hobby--Eberly Telescope (HET) and the McDonald 2.7m Harlan--Smith Telescope 
are shown as filled circles (see Table 2 for photometric information). 

\subsection{Observations}

Follow-up spectroscopy of 9 white dwarf candidates in the Spaghetti Survey Field was
obtained in April and May 2002 using the HET and in February 2003 using the 2.7m Harlan--Smith Telescope. 
We used the HET equipped with the Marcario Low Resolution Spectrograph (LRS) to obtain low 
resolution spectroscopy of four cool white dwarf
candidates. Grism 1 with a 2" slit produced spectra with a resolution of 16 \AA\ over the 
range 4000 -- 10000 \AA. Spectroscopy for four additional stars was obtained at
the McDonald 2.7m Telescope with the Imaging Grism Instrument (IGI) and TK4 camera using the
holographic grating, which produced spectra with a resolution of 12 \AA\ over the range 4000 --
8000 \AA. A spectrophotometric standard star was observed each night for flux calibration.
Ne--Cd calibration lamp exposures were taken after each observation
with the HET, and Ne--Ar lamp calibrations were taken at the beginning of the night for the 2.7m
observations. The data were reduced using standard IRAF\footnote{IRAF is distributed by the National Optical Astronomy Observatory, which is operated by the Association of Universities for Research in Astronomy (AURA), Inc., under cooperative agreement with the National Science Foundation.} routines.

\subsection{Results}

The observed spectra for selected white dwarf candidates from the HET and the 2.7m and the fitted template spectra (shown in red) are shown in Figure 3a and 3b,
respectively. Spectra are ordered by $g-r$ color.  We have used Pickles (1998) template spectra
to classify the observed spectra qualitatively. The object numbers, coordinates and the assigned spectral classifications are shown on
the lower right corner of the figures.
Two of the objects observed at the HET, SDSS J114149.41$-$001140.4 and SDSS
J120709.16$-$011247.2, show blue excesses.
SDSS J114149.41$-$001140.4 also shows strong H$_{\beta}$ and H$_{\gamma}$ lines.
Therefore we classify these stars as white dwarf + late type star spectroscopic binaries.
Figure 3a and 3b show that none of the observed white dwarf candidates are actually single white dwarfs.
This discovery contradicted the expected yield of the DDO51 filter which led us to reconsider our strategy for
using this filter. We then pursued a reverse approach which
is described in the next section.

\section{Reverse Approach: Spectroscopy to Photometry}

Follow--up spectroscopy of photometrically selected cool white dwarf candidates resulted in the discovery of 
subdwarf stars and unusual binaries
instead of cool white dwarfs. In order to test the effectiveness of the filter in distinguishing
cool white dwarfs from
subdwarf stars, we decided to observe known cool white dwarfs with the DDO51 filter.

\subsection{Observations and Results}
DDO51, r, and z band photometry of 30 known cool white dwarfs with temperatures ranging from 10000 K down to very
cool temperatures ($T_{\rm eff}\leq3500$ K) was obtained at the CTIO 4m--Blanco Telescope and 
Kitt Peak 4m--Mayall Telescope
equipped with the $8k \times 8k$ MOSAIC Imager in November 2002 and June 2003, respectively. The MOSAIC Imager when
used with these 4m telescopes provides a 35' $\times$ 35' field of view. The CCD images were processed with the standard
procedures in the MSCRED package in IRAF v2.12. We adopted the reduction procedures used by the NOAO Deep Wide--Field
Survey Team\footnote{MOSAIC reduction procedures can be found at http://www.noao.edu/noao/noaodeep/ReductionOpt/frames.html}. Images were bias subtracted and flat--fielded using dome flats and sky flats.
The z band images were corrected for fringing using the fringing templates derived from the sky flats.
We matched star positions in
the fields with positions from the USNO--A V2.0 Catalog (Monet et al. 1998) to obtain a plate solution for each
CCD. The RMS differences between observed source positions and the USNO--A V2.0 Catalog were less than 0.4 arcseconds.
Using the derived astrometric solutions, images were projected to the same uniform (linear, 0.258 arcsecond per pixel)
scale.
 
Source identifications were performed on the projected images using the SExtractor package (Bertin \& Arnouts 1996) v2.1.6.
The main motivation for using the SExtractor package was its morphological classification. SExtractor uses a neural
network to classify objects as stars (stellarity$=$1) or galaxies (stellarity$=$0). Stellarity is
a continuous variable that can take any value from 0 to 1.
A comparison of stellarity indices with magnitudes show that
objects with stellarity index $\geq$ 0.8 have reliable classification as stars.
We tested the SExtractor and the IRAF routines WPHOT and PHOT to perform photometry on our images. WPHOT with a gaussian weighting scheme gave the
best results for fainter objects, therefore we adopted WPHOT photometry.
We selected all
objects with a stellarity index larger than 0.8 and with photometric errors less than 0.1 mag for our analysis. Only those objects
detected in each filter that matched up to within 0.5 arcsec or better in each coordinate are included in our final
catalogs.

Most of our observations were obtained under photometric conditions. Since we are mainly
interested in the differential photometry
between white dwarfs and the rest of the field stars, data from non--photometric nights are also useful.
We have cross-correlated color--color diagrams for each field with the data from the Spaghetti Survey and matched
the observed field star sequences to remove any photometric offsets from non--photometric observing conditions. Figure 4 shows the color--color diagram for 30 known white dwarfs and surrounding field stars. Field stars from  
the Spaghetti Survey and our study are shown as black dots.
A typical errorbar for the field stars is shown in the lower left corner of the figure.
A good match between our data and the Spaghetti Survey data is apparent in this figure.
Known white dwarfs are shown as filled circles.
Our synthetic photometry of white dwarf model atmospheres
(Saumon \& Jacobson 1999; D. Saumon, private communication) for pure H (solid line)
and pure He (dashed line) white
dwarfs with 7000 $\gtrsim T_{\rm eff} \gtrsim$ 3000 K and a blackbody (dotted line) are also shown.
Dashed--dotted lines represent
mixed H/He atmospheres for 3500 K and 3000 K white dwarfs with different compositions (log [N(He)/N(H)]$=-1$ through 6).

Temperatures and colors for the observed white dwarfs are given in Table 3. The observed white dwarf sequence
is in agreement with our follow--up spectroscopy, and both demonstrate that white dwarfs are much closer to (and more
blended with) the main sequence stars than previously predicted. Cool white dwarfs occupy a region running
from the center of the field star locus ($r-DDO=-0.1$, $r-z=0.05$) for $T_{\rm eff}\sim$
7000 K to the red edge of the field star locus ($r-DDO=-0.75$, $r-z=0.65$).

\section{Discussion}

Our observations demonstrate that the narrow-band DDO51 filter, centered on the Mg band,
is not as effective at separating white dwarfs from subdwarfs
as we expected.
White dwarfs with temperatures between
7000 K and 5000 K ($-0.10\geq r-DDO\geq-0.45$, $0.05\leq r-z\leq0.35$) are
photometrically indistinguishable from observed field stars.
Using template spectra from the Pickles (1998) library, we
have measured the equivalent width of the Mg/MgH feature in
main sequence stars.
Mg absorption becomes strong enough to affect the photometry
in K0 ($T_{\rm eff}\sim$ 5000 K) and later type stars (see Figure 5).
Due to the spread in colors and weak Mg absorption in
the F--G type stars,
white dwarfs with 7000 K$\geq T_{\rm eff}\geq$ 5000 K have
similar colors to F--G stars.

White dwarfs with temperatures in the range 5000 -- 3500 K ($-0.45\geq r-DDO\geq-0.80$,
$0.35\leq r-z\leq0.65$) lie just above the edge of the observed field star sequence.
Until recently, cool white dwarfs were thought to have spectral energy
distributions similar to blackbodies. In fact, this is why
Claver (1995) suggested that a narrow-band filter centered on the MgH feature
would place cool white dwarfs above the observed field star sequence; the DDO51 filter would
separate blackbodies from subdwarfs (see Figure 1 and Figure 4).
Although subdwarfs have strong MgH absorption in this temperature
range (Figure 5) and they deviate from blackbodies, observed
white dwarfs deviate from blackbodies, too.
The effects of collision induced absorption (CIA) due to molecular hydrogen are expected to be significant below 5000 K (Hansen et al. 1998; Saumon \& Jacobson 1999).
Figure 4 shows that in these colors, there 
are no pure H white dwarfs with $T_{\rm eff}\leq$5000 K and the observed
white dwarf sequence
actually continues along between the pure H and the pure He models.
This is also seen in the $B-V$ vs. $V-K$ color--color diagrams of Bergeron,
Ruiz \& Leggett (1997) and Bergeron, Leggett \& Ruiz (2001)
which implies that either all cool white dwarfs have mixed H/He composition,
the calculated CIA opacities are incorrect, or there are
other neglected physical effects.
We note that Bergeron \& Leggett (2002) found that all white dwarfs
cooler than 4000 K have mixed H/He atmospheres.
Even if the white dwarf model atmospheres and the CIA opacities are right, the question
of why we still have
not found a pure H white dwarf that shows CIA remains to be answered.
A possible explanation for the lack of discovery of such objects may simply be the finite
age of the Galactic disk; pure H white dwarfs have not yet cooled enough to show strong CIA
absorption.

The effective temperature range between 5000 and 3500 K is the most important
regime for white dwarf luminosity function studies since it defines
the turn--off of the white dwarf luminosity function, hence the age of the observed population.
A single slit spectrograph would not be efficient in finding those objects preselected by the
DDO51 photometry technique, but a wide field multi--object
spectrograph, e.g. Hectospec (Fabricant, Hertz, \& Szentgyorgyi 1994) with 300 fibers
on the converted
Multiple Mirror Telescope, might be used productively to
carve out regions from the $r-DDO51$ vs. $r-z$ color--color diagram
to find cool white dwarfs in this range.
Figure 4 shows a possible search box (shown in green) for cool white dwarfs.
For a one square degree field at a Galactic latitude $l=38$, the box
includes 234 stars down to $r=21.5$.
Using the Liebert, Dahn \& Monet (1988) white dwarf luminosity function
and a disk scale height of 250 pc,
we expect to find one cool white dwarf per square degree in the
search box (5000 K$\geq T_{\rm eff}\geq$
3500 K). In other words, the average pointing with the MMT + Hectospec
should yield a cool white dwarf. 
The above field has 660 stars in the color range $-0.80\leq r-DDO\leq-0.45$ and 1014 stars
in the range $0.35\geq r-z\geq0.65$.
Even though the DDO51 filter technique is not as efficient as expected, it rejects at least 65\% of main sequence
stars in this temperature range. Therefore, it is $\sim$3 times more efficient
than purely spectroscopic (i.e. no prior photometry) surveys.
The DDO51 filter is widely used to identify halo stars and to distinguish between giants and dwarfs
(Morrison et al. 2001). Thus, as a byproduct, DDO51 photometry from the Spaghetti Survey
and similar surveys can be used
to identify cool white dwarf candidates for follow-up spectroscopy.

Four ultra cool white dwarfs (CE51, LHS3250, SDSS J133739.40+000142, and LHS1402) lie to the left of the field
stars and are clearly separated from the observed sequence of stars due to their depressed near--infrared colors which is thought to be the result of CIA absorption.
DDO51 filter photometry is not necessary for finding ultra cool white dwarfs since
these stars have broad molecular features and they can be found using broad-band
photometry, e.g. in the Sloan Digital Sky Survey.
On the other hand, it can help identify the elusive He-rich ultra cool
white dwarfs because
they approximate a blackbody spectral energy distribution.
The four ultra cool white dwarfs in Figure 4 appear to form a sequence.
Ruiz \& Bergeron (2001) find an H--dominated atmosphere solution with a temperature estimate of 2730 K for CE51, though infrared photometry is needed to determine the temperature of this star reliably.
Simply based on Figure 4, CE51 is more readily explained as 
a $\sim$ 3200 K white dwarf of mixed composition.
Bergeron \& Leggett (2002) tried to fit the spectra for LHS3250 and SDSS J133739.40+000142, and found that they are
inconsistent with being pure H atmosphere stars. Mixed H/He atmosphere composition is predicted by Bergeron \& Leggett (2002),
yet the overall shape of the spectra cannot be fitted with the current model atmospheres.
Farihi (2004) has found yet another cool white dwarf, GD392B, consistent with a mixed H/He
atmosphere. Estimated tangential velocities for the four ultra cool white dwarfs and GD392B are
consistent
with them being disk objects and their excess luminosity may be explained
if they are low--mass white dwarfs or unusual spectroscopic binaries (Ruiz \& Bergeron 2001;
Harris et al. 2001; Bergeron 2003; Farihi 2004).

Mixed atmosphere white dwarfs cool faster than their pure--H counterparts,
therefore they are not the defining stars for the age estimates for the Galactic disk, unless
no pure--H atmosphere white dwarfs exist below 4000 K. Although the four
ultra cool white dwarfs appear to form a sequence in the $r-DDO51$ vs. $r-z$ color--color diagram, we do not understand their nature at this time.
Also, they are not on
the theoretically predicted blue hook for cool hydrogen-rich white dwarfs. Our observations demonstrate that
optical colors should not be used to estimate the temperatures and ages of ultra cool
white dwarfs since
current model atmospheres are not fully capable of explaining their observed colors.

Mg/MgH and $CaH+TiO$ are the most prominent features in the optical spectra of subdwarf stars.
In addition to the DDO51 filter, we have also investigated the use of an intermediate-band
filter centered on the $CaH+TiO$ band at $\sim$6850\AA\ (Claver 1995)
to test whether it can be used to identify white dwarfs. Equivalent width measurements
of this band using the Pickles (1998) template
spectra are shown in Figure 5. $CaH+TiO$ absorption becomes stronge in M0 and later type stars.
White dwarfs in this temperature range show depressed infared colors due to CIA
if they have pure--H or mixed H/He atmospheres, and they can be
identified by using the DDO51 filter if they have pure--He atmospheres (true--blackbody).
The CIA exhibited by ultra cool white dwarfs is extremely broad-band
and monotonically varies throughout the red-infrared region, whereas the
CaH/TiO band is very narrowly confined in wavelength. Thus, the $CaH+TiO$ filter, if ratioed
with another nearby pseudocontinuum filter, could show a much
stronger dependency on temperature and metallicity in main sequence and
subdwarf stars than it does in ultra cool white dwarfs.
Therefore, the $CaH+TiO$ filter and/or $JHK$ infrared photometry may be useful for the identification
of cool hydrogen-rich or mixed atmosphere white dwarfs, though
broad-band photometry surveys are also successful in
finding ultra cool white dwarfs (e.g. Sloan Digital Sky Survey; Harris et al. 2001; Gates et al. 2004).

\acknowledgements
We thank Jennifer Claver for useful discussions on reducing MOSAIC data and
the NOAO Deep Wide--Field Survey Team for making their reduction procedures available online.
We also thank Didier Saumon for making his cool white dwarf model atmospheres available
to us and for careful reading of this manuscript. We are grateful to Ed Olsewzki for making his DDO51 photometry data available to us.
This material is based upon work supported by the National Science Foundation
under Grant No. 0307315.
The Hobby-Eberly Telescope (HET) is a joint project of the University of Texas at Austin, the Pennsylvania State University, Stanford University, Ludwig-Maximilians-Universit\"at M\"unchen, and Georg-August-Universit\"at G\"ottingen. The HET is named in honor of its principal benefactors, William P. Hobby and Robert E. Eberly. The Marcario Low Resolution Spectrograph is named for Mike Marcario of High Lonesome Optics who fabricated several optics for the instrument but died before its completion. The LRS is a joint project of the Hobby-Eberly Telescope partnership and the Instituto de Astronomía de la Universidad Nacional Autonoma de M\'{e}xico.

\clearpage
\begin{deluxetable}{llrrrrrrrrrrrrl}
\tabletypesize{\scriptsize}
\rotate
\tablecolumns{15}
\tablewidth{0pt}
\tablecaption{Objects with SDSS Spectroscopy}
\tablehead{
\colhead{Object}&
\colhead{Plate MJD Fiber}&
\colhead{ddo}&
\colhead{u}&
\colhead{g}&
\colhead{r}&
\colhead{i}&
\colhead{z}&
\colhead{$\sigma_{ddo}$}&
\colhead{$\sigma_{u}$}&
\colhead{$\sigma_{g}$}&
\colhead{$\sigma_{r}$}&
\colhead{$\sigma_{i}$}&
\colhead{$\sigma_{z}$}&
\colhead{Type}
}
\startdata
SDSS J101748.90$-$003124.5& 271 51883 166& 18.68& 19.30& 18.86& 18.84& 18.85& 18.65& 0.01& 0.03& 0.01& 0.01& 0.01& 0.03& QSO\\
SDSS J101807.04$-$002003.3& 271 51883 174& 20.14& 20.99& 19.98& 19.91& 19.97& 19.49& 0.02& 0.12& 0.02& 0.03& 0.04& 0.08& QSO\\
SDSS J101741.70$-$002934.1& 271 51883 181& 16.34& 16.56& 16.28& 16.72& 16.88& 17.17& 0.01& 0.01& 0.01& 0.01& 0.01& 0.02& WD\\
SDSS J101651.74$-$003347.0& 271 51883 226& 19.14& 18.93& 19.01& 18.95& 18.65& 18.64& 0.01& 0.02& 0.02& 0.01& 0.01& 0.04& QSO\\
SDSS J105907.68+010303.5& 277 51908 361& 19.26& 19.20& 19.17& 18.88& 18.78& 18.83& 0.01& 0.03& 0.01& 0.02& 0.02& 0.04& QSO\\
SDSS J105934.61+011112.1& 277 51908 377& 17.81& 18.82& 17.86& 17.50& 17.38& 17.34& 0.01& 0.02& 0.01& 0.03& 0.01& 0.02& Star\\
SDSS J110015.66+010740.5& 277 51908 414& 18.80& 19.09& 18.74& 18.99& 19.30& 19.57& 0.01& 0.02& 0.01& 0.02& 0.03& 0.08& WD\\
SDSS J110010.68+010328.3& 277 51908 416& 16.89& 18.97& 17.63& 17.40& 17.31& 17.30& 0.01& 0.02& 0.01& 0.02& 0.02& 0.02& Star\\
SDSS J112941.64+000545.3& 281 51614 117& 18.60& 19.63& 18.61& 18.57& 18.53& 18.61& 0.01& 0.04& 0.02& 0.02& 0.03& 0.06& Star\\
SDSS J112837.56$-$000112.6& 281 51614 142& 19.05& 19.57& 18.35& 18.32& 18.38& 18.34& 0.01& 0.03& 0.01& 0.01& 0.01& 0.03& Star\\
SDSS J112839.72+002644.3& 281 51614 485& 18.11& 19.35& 18.19& 17.98& 17.99& 18.01& 0.01& 0.04& 0.03& 0.01& 0.02& 0.03& Star\\
SDSS J113015.48+002843.9& 281 51614 551& 17.74& 17.71& 17.68& 17.72& 17.93& 17.90& 0.01& 0.01& 0.02& 0.02& 0.02& 0.02& QSO\\
SDSS J113026.84+002649.1& 282 51658 348& 20.20& 20.01& 19.99& 20.09& 19.83& 19.77& 0.02& 0.05& 0.03& 0.03& 0.04& 0.11& QSO\\
SDSS J113031.57+002033.4& 282 51658 353& 17.95& 18.08& 18.04& 17.87& 17.88& 17.84& 0.01& 0.01& 0.02& 0.01& 0.02& 0.02& QSO\\
SDSS J113003.26+000332.4& 282 51658 356& 19.63& 19.64& 19.53& 19.55& 19.37& 19.23& 0.01& 0.04& 0.02& 0.03& 0.04& 0.09& QSO\\
SDSS J113009.77+001737.6& 282 51658 357& 19.70& 20.89& 19.81& 19.80& 19.40& 19.36& 0.02& 0.06& 0.03& 0.03& 0.02& 0.06& Star\\
SDSS J114311.23$-$002133.0& 283 51959 224& 18.99& 18.88& 18.92& 18.83& 18.72& 18.56& 0.01& 0.02& 0.02& 0.02& 0.03& 0.04& QSO\\
SDSS J114321.76$-$002941.5& 283 51959 229& 18.81& 19.32& 19.10& 19.11& 19.26& 18.96& 0.01& 0.03& 0.01& 0.02& 0.02& 0.04& QSO\\
SDSS J114318.48$-$002254.9& 283 51959 237& 18.59& 19.57& 18.63& 18.45& 18.46& 18.42& 0.01& 0.03& 0.02& 0.02& 0.03& 0.03& Star\\
SDSS J114210.47$-$002013.0& 283 51959 238& 19.08& 19.36& 19.16& 18.89& 18.87& 18.76& 0.01& 0.03& 0.02& 0.03& 0.05& 0.04& QSO\\
SDSS J114137.14$-$002729.8& 283 51959 262& 18.61& 18.94& 18.75& 18.59& 18.42& 18.18& 0.01& 0.02& 0.01& 0.02& 0.01& 0.03& QSO\\
SDSS J114151.31$-$000729.7& 283 51959 398& 17.10& 18.05& 17.17& 16.90& 16.82& 16.81& 0.01& 0.01& 0.01& 0.01& 0.02& 0.02& Star\\
SDSS J114259.30$-$000156.5& 283 51959 435& 19.47& 19.96& 19.50& 19.18& 18.93& 18.69& 0.02& 0.04& 0.01& 0.02& 0.02& 0.03& QSO\\
SDSS J120637.75$-$011246.6& 286 51999 81& 17.29& 18.24& 17.34& 17.12& 17.00& 16.98& 0.01& 0.02& 0.01& 0.02& 0.01& 0.02& Star\\
SDSS J120631.48$-$010801.8& 286 51999 85& 19.37& 20.38& 19.44& 19.12& 19.02& 18.87& 0.01& 0.07& 0.02& 0.02& 0.03& 0.06& Star\\
SDSS J120708.35$-$010002.5& 286 51999 86& 19.07& 19.12& 19.09& 18.94& 18.59& 18.50& 0.01& 0.03& 0.02& 0.01& 0.02& 0.03& QSO\\
SDSS J123056.59$-$005306.4& 289 51990 17& 18.89& 19.26& 18.87& 18.73& 18.72& 18.46& 0.01& 0.03& 0.02& 0.01& 0.01& 0.03& QSO\\
SDSS J123213.28$-$003106.3& 289 51990 25& 19.08& 20.22& 19.07& 19.15& 19.21& 19.28& 0.02& 0.04& 0.02& 0.01& 0.02& 0.05& Star\\
SDSS J123051.62$-$004437.1& 290 51941 284& 18.00& 19.00& 17.84& 17.97& 18.03& 18.14& 0.01& 0.02& 0.02& 0.02& 0.02& 0.03& Star\\
SDSS J123149.22$-$005550.4& 290 51941 292& 17.49& 18.45& 17.46& 17.43& 17.48& 17.53& 0.01& 0.01& 0.02& 0.01& 0.01& 0.01& Star\\
SDSS J123140.81$-$004435.1& 290 51941 293& 18.71& 19.68& 18.75& 18.62& 18.58& 18.56& 0.01& 0.03& 0.02& 0.02& 0.02& 0.04& Star\\
SDSS J123027.54$-$003633.3& 290 51941 319& 16.88& 17.84& 16.79& 17.07& 17.29& 17.37& 0.01& 0.01& 0.02& 0.01& 0.02& 0.01& Star\\
SDSS J131941.10$-$004340.6& 296 51984 184& 20.46& 23.35& 20.85& 19.27& 17.21& 16.03& 0.03& 0.68& 0.04& 0.02& 0.01& 0.01& Star\\
SDSS J131938.76$-$004940.0& 296 51984 189& 17.59& 17.73& 17.54& 17.48& 17.50& 17.40& 0.01& 0.02& 0.02& 0.01& 0.01& 0.02& QSO\\
SDSS J143143.80$-$005011.4& 306 51637 129& 18.05& 18.31& 18.09& 17.85& 17.81& 17.89& 0.01& 0.02& 0.02& 0.02& 0.02& 0.02& QSO\\
SDSS J143158.36$-$004303.9& 306 51637 138& 19.14& 19.06& 18.99& 18.80& 18.74& 18.88& 0.01& 0.02& 0.03& 0.02& 0.02& 0.03& QSO\\
SDSS J143153.06$-$002824.3& 306 51637 194& 17.82& 17.29& 17.66& 18.21& 18.91& 18.85& 0.01& 0.02& 0.02& 0.02& 0.03& 0.04& SDO\\
SDSS J143037.11$-$004748.9& 306 51637 255& 16.98& 17.92& 17.02& 16.89& 16.88& 16.89& 0.01& 0.02& 0.01& 0.01& 0.02& 0.01& Star\\
SDSS J145321.76+010130.6& 538 52029 212& 19.95& 20.53& 20.10& 20.15& 19.96& 19.68& 0.02& 0.06& 0.02& 0.03& 0.04& 0.08& QSO\\
SDSS J145244.19+010954.9& 538 52029 251& 19.97& 20.21& 20.21& 19.90& 19.88& 19.56& 0.02& 0.05& 0.03& 0.03& 0.03& 0.08& QSO\\
\enddata
\end{deluxetable}

\clearpage
\begin{deluxetable}{clrrrrrrrrrrrr}
\rotate
\tablecolumns{14}
\tablewidth{0pt}
\tablecaption{Objects with HET + McDonald 2.7m Spectroscopy}
\tablehead{
\colhead{No}&
\colhead{Object}&
\colhead{ddo}&
\colhead{u}&
\colhead{g}&
\colhead{r}&
\colhead{i}&
\colhead{z}&
\colhead{$\sigma_{ddo}$}&
\colhead{$\sigma_{u}$}&
\colhead{$\sigma_{g}$}&
\colhead{$\sigma_{r}$}&
\colhead{$\sigma_{i}$}&
\colhead{$\sigma_{z}$}
}
\startdata
1& SDSS J114149.41$-$001140.4& 19.67& 20.49& 19.77& 19.14& 18.05& 17.32& 0.01& 0.05& 0.02& 0.02& 0.02& 0.02\\
2& SDSS J120650.72$-$010519.1& 20.29& 22.44& 20.81& 19.44& 17.87& 17.11& 0.02& 0.43& 0.06& 0.05& 0.02& 0.03\\
3& SDSS J120651.91$-$010435.2& 19.31& 21.30& 19.38& 18.59& 18.32& 18.14& 0.01& 0.16& 0.01& 0.02& 0.02& 0.03\\
4& SDSS J120709.16$-$011247.2& 16.84& 18.12& 16.90& 16.34& 15.70& 15.33& 0.02& 0.01& 0.01& 0.02& 0.01& 0.02\\
5& SDSS J120741.55$-$010630.9& 19.89& 22.70& 20.07& 19.62& 18.39& 18.17& 0.02& 1.17& 0.15& 0.19& 0.08& 0.09\\
6& SDSS J123202.54$-$010232.1& 20.33& 22.88& 20.59& 19.25& 17.94& 17.25& 0.04& 0.98& 0.03& 0.02& 0.01& 0.01\\
7& SDSS J123208.81$-$010230.5& 20.69& 22.52& 20.68& 19.82& 19.37& 19.03& 0.04& 0.31& 0.03& 0.02& 0.02& 0.04\\
8& SDSS J131833.56$-$004448.4& 16.59& 18.76& 16.80& 16.11& 15.68& 15.41& 0.01& 0.02& 0.02& 0.01& 0.02& 0.01\\
9& SDSS J143044.16$-$002853.1& 16.99& 18.74& 17.09& 16.53& 16.35& 16.63& 0.01& 0.03& 0.01& 0.01& 0.02& 0.02\\
\enddata
\end{deluxetable}

\clearpage
\begin{deluxetable}{lrrrrrlc}
\tabletypesize{\scriptsize}
\tablecolumns{8}
\tablewidth{0pt}
\tablecaption{DDO51 photometry for previously identified white dwarfs}
\tablehead{
\colhead{Object}&
\colhead{R}&
\colhead{$r-z$}&
\colhead{$\sigma_{r-z}$}&
\colhead{$r-ddo$}&
\colhead{$\sigma_{r-ddo}$}&
\colhead{$T_{\rm eff}$(K)}&
\colhead{Notes}
}
\startdata
WD2323+157&                     15.04& $-$0.31& 0.01&  0.19& 0.01& 10170& 1\\
LTT 9491&                       14.07& $-$0.37& 0.01&  0.19& 0.01& \nodata & 2\\
CE157&                          15.64&  0.03& 0.03& $-$0.07& 0.02& 7000& 3\\
WD1325+581&                     16.42&  0.07& 0.01& $-$0.15& 0.01& 6810& 4\\
WD1633+572&                     14.68&  0.08& 0.01& $-$0.16& 0.01& 6180& 4\\
WD2107-216&                     16.45&  0.09& 0.01& $-$0.21& 0.01& 5830& 4\\
WD2347+292&                     15.41&  0.13& 0.01& $-$0.14& 0.01& 5810& 4\\
WD0752$-$676&                     13.58&  0.09& 0.01& $-$0.22& 0.01& 5730& 4\\
CE162&                          16.69&  0.12& 0.01& $-$0.21& 0.01& 5730& 3\\
WD1257+037&                     15.46&  0.21& 0.01& $-$0.22& 0.01& 5590& 4\\
WD2248+293&                     15.14&  0.19& 0.01& $-$0.21& 0.01& 5580& 4\\
WD0121+401&                     16.67&  0.28& 0.01& $-$0.31& 0.01& 5340& 4\\
WD1334+039&                     14.12&  0.36& 0.01& $-$0.43& 0.01& 5030& 4\\
WD2002$-$110&                     16.36&  0.29& 0.01& $-$0.38& 0.01& 4800& 4\\
WD0045$-$061&                     17.70&  0.32& 0.01& $-$0.40& 0.01& \nodata& 5\\
WD1820+609&                     15.15&  0.37& 0.01& $-$0.45& 0.01& 4780& 4\\
LHS 542&                        17.53&  0.44& 0.04& $-$0.57& 0.01& 4720& 4\\
WD1108+207&                     17.17&  0.45& 0.01& $-$0.50& 0.01& 4650& 4\\
WD1345+238&                     15.12&  0.45& 0.01& $-$0.57& 0.01& 4590& 4\\
WD2251$-$070&                     15.10&  0.46& 0.01& $-$0.54& 0.01& 4580& 4\\
CE40&                           18.82&  0.50& 0.01& $-$0.63& 0.01& 4580& 3\\
CE142&                          18.59&  0.55& 0.02& $-$0.70& 0.02& 4390& 3\\
CE16&                           17.59&  0.64& 0.01& $-$0.78& 0.01& 4330& 3\\
WD1300+263&                     18.09&  0.56& 0.01& $-$0.67& 0.01& 4320& 4\\
WD1247+550&                     17.03&  0.66& 0.01& $-$0.75& 0.01& 4050& 4\\
F351$-$50&                        18.37&  0.56& 0.01& $-$0.80& 0.01& 3500:& 5\\
CE51&                           17.50& $-$0.36& 0.01& $-$0.55& 0.01& 2730:& 3\\
LHS 3250&                       17.87& $-$0.69& 0.03& $-$0.26& 0.01& \nodata & 6\\
SDSS J133739.40+000142.8&       19.16& $-$0.82& 0.11& $-$0.20& 0.02& \nodata  & 6\\
LHS 1402&                       17.86& $-$0.96& 0.02&  0.03& 0.01& \nodata & 5\\
\enddata
\tablecomments{
(1) R magnitude and $T_{\rm eff}$ from Bergeron, Ruiz \& Leggett 1997; (2) Spectral Type DBQA5 (Wesemeal et al. 1993); (3) R magnitude and $T_{\rm eff}$ from Ruiz \& Bergeron 2001; (4) R magnitude and $T_{\rm eff}$ from Bergeron, Leggett \& Ruiz 2001; (5) R$_{59F}$ magnitude and/or $T_{\rm eff}$ from Oppenheimer et al. 2001; (6) SDSS r magnitude
}
\end{deluxetable}

\clearpage
\figcaption[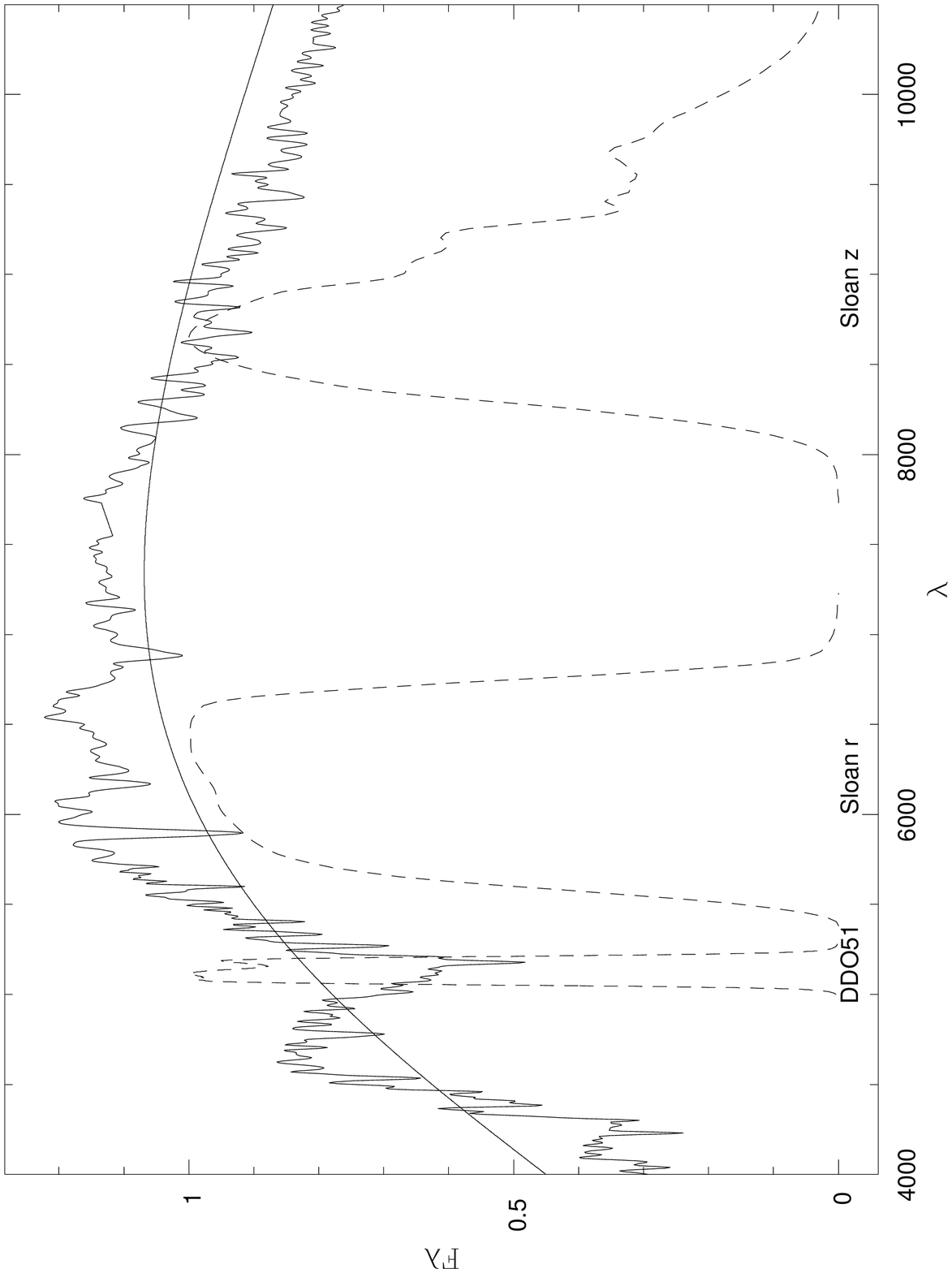]{
Template spectra for a K5V star (Pickles 1998) and a 5000 K blackbody. The tracing for the DDO51,
r, and z
filters are shown as dotted--lines. The greatest difference between the blackbody and the
template stellar spectra,
Mg absorption, is apparent in this figure.
\label{fig1}}

\figcaption[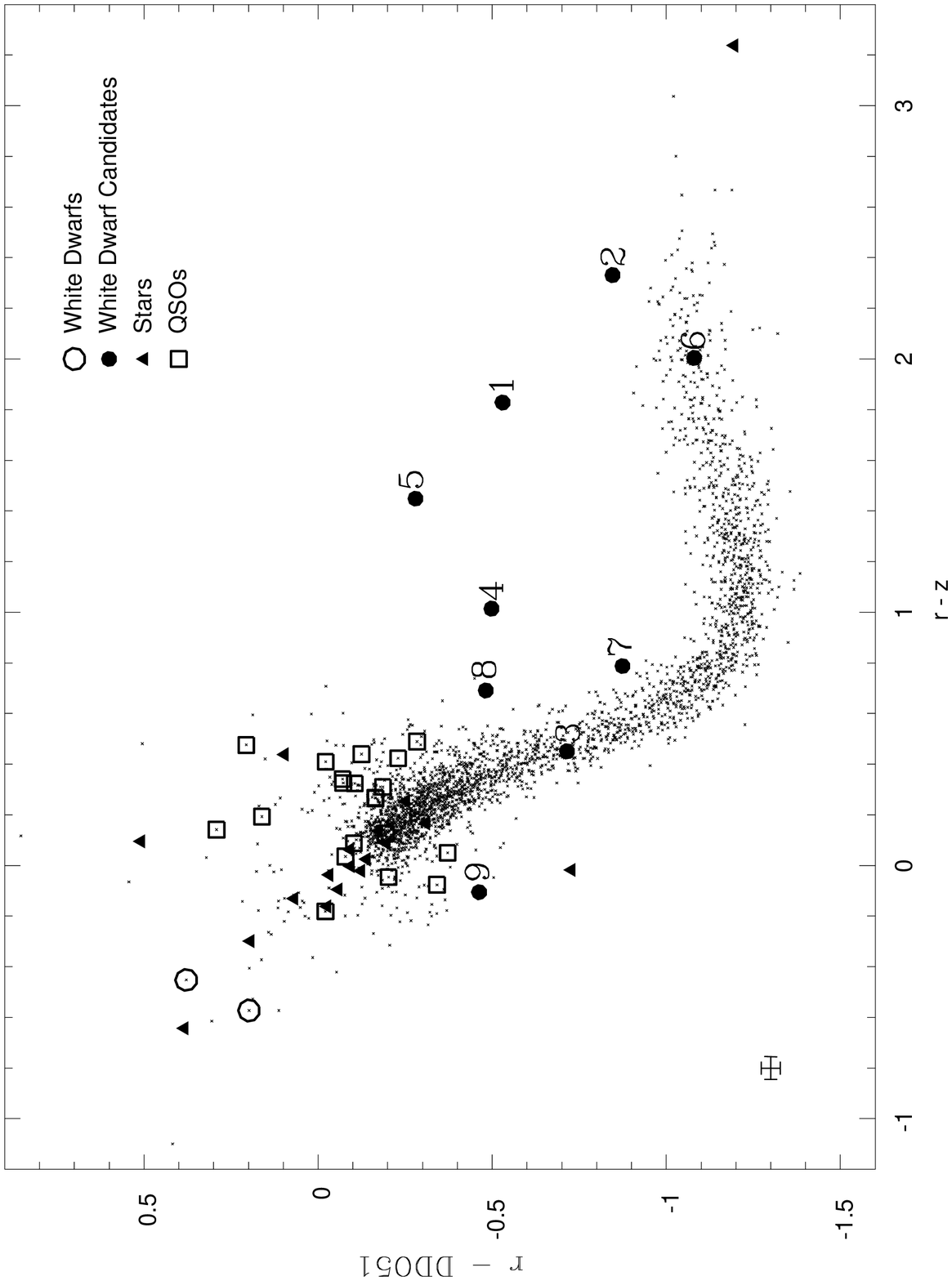]{
$r-DDO51$ vs. $r-z$ color--color diagram of a 2 square degree field from the Spaghetti Survey which
overlaps with the Sloan Digital Sky Survey fields. Two hot, spectroscopically identified white
dwarfs are shown as open circles. Quasars and stars (which are not white dwarfs) are shown as
open squares and filled triangles, respectively. Cool white dwarf candidates selected for follow-up
spectroscopy are shown as filled circles.
We note that \# 3 and \#6 were not selected as cool white dwarf candidates, but they happened to be
positioned on the slit during our observations of cool white dwarf candidates.
\label{fig2}}

\figcaption[kilic.fig3.ps]{
Optical spectra for the white dwarf candidates observed at the HET (Figure 3a) and the 2.7m
(Figure 3b). Template spectra (Pickles 1998) used to classify these objects are shown in red.
Object numbers, coordinates and spectral types are shown in the lower right corner.
Note that the feature at 7600 \AA\ is telluric and the emission features in \# 1 at $\sim$ 7700 \AA\
and in \# 2 at $\sim$ 4300 \AA\ are due to cosmic ray hits.
\label{fig3}}

\figcaption[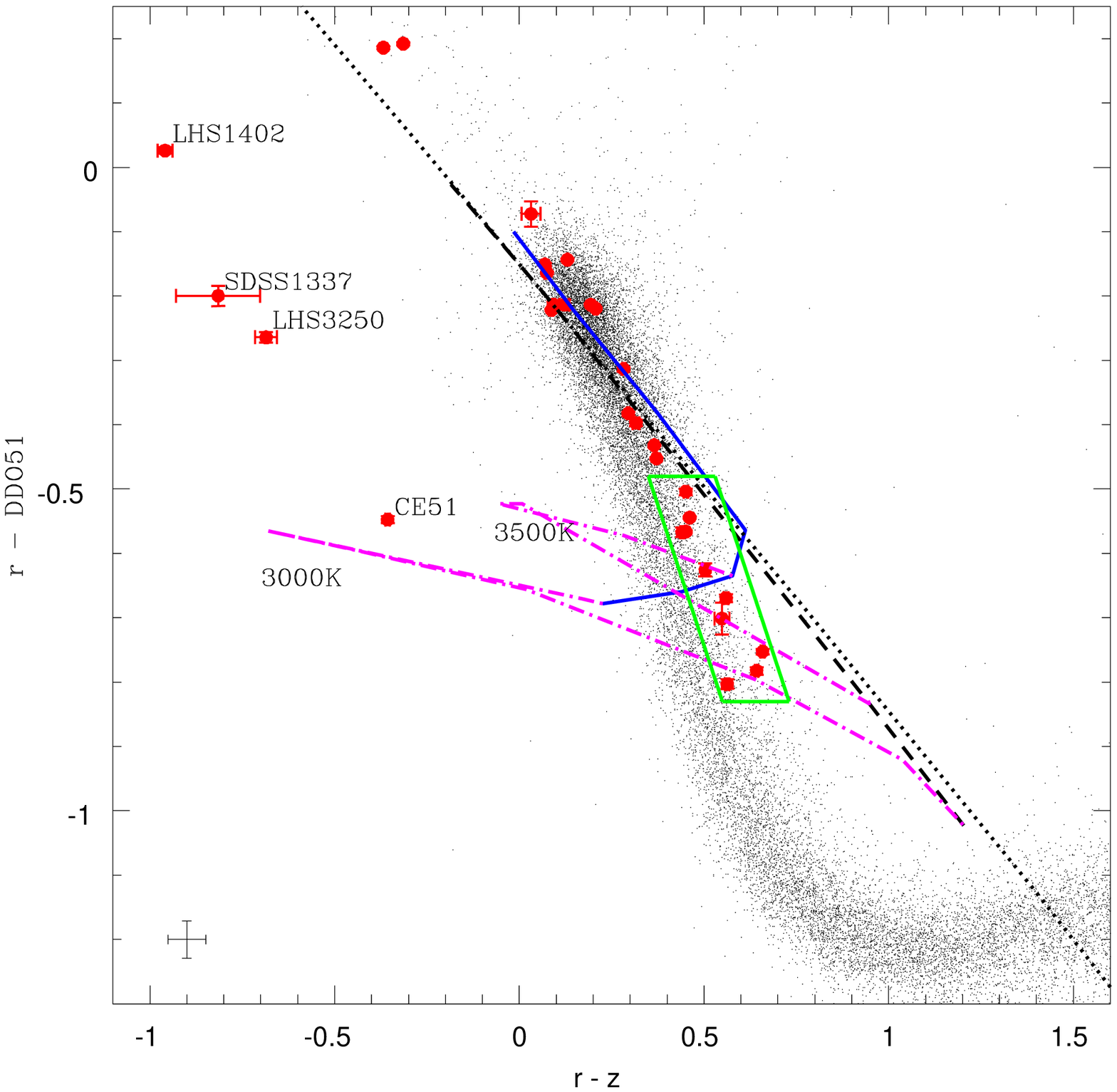]{
$r-DDO51$ vs. $r-z$ color--color diagram for 30 known white dwarfs (filled circles) and the
surrounding field stars (black dots).
A typical errorbar for the field stars is shown in the lower left corner.
Our synthetic photometry of pure H and pure He white dwarf models with 7000 $\gtrsim T_{\rm eff} \gtrsim$ 3000 K are shown as solid and dashed lines, respectively. Dashed--dotted lines represent mixed atmospheres for 3500 K and 3000 K white dwarfs with different compositions and a blackbody is also shown as a dotted line.
The green box marks a possible search region for cool white dwarfs.
\label{fig4}}

\figcaption[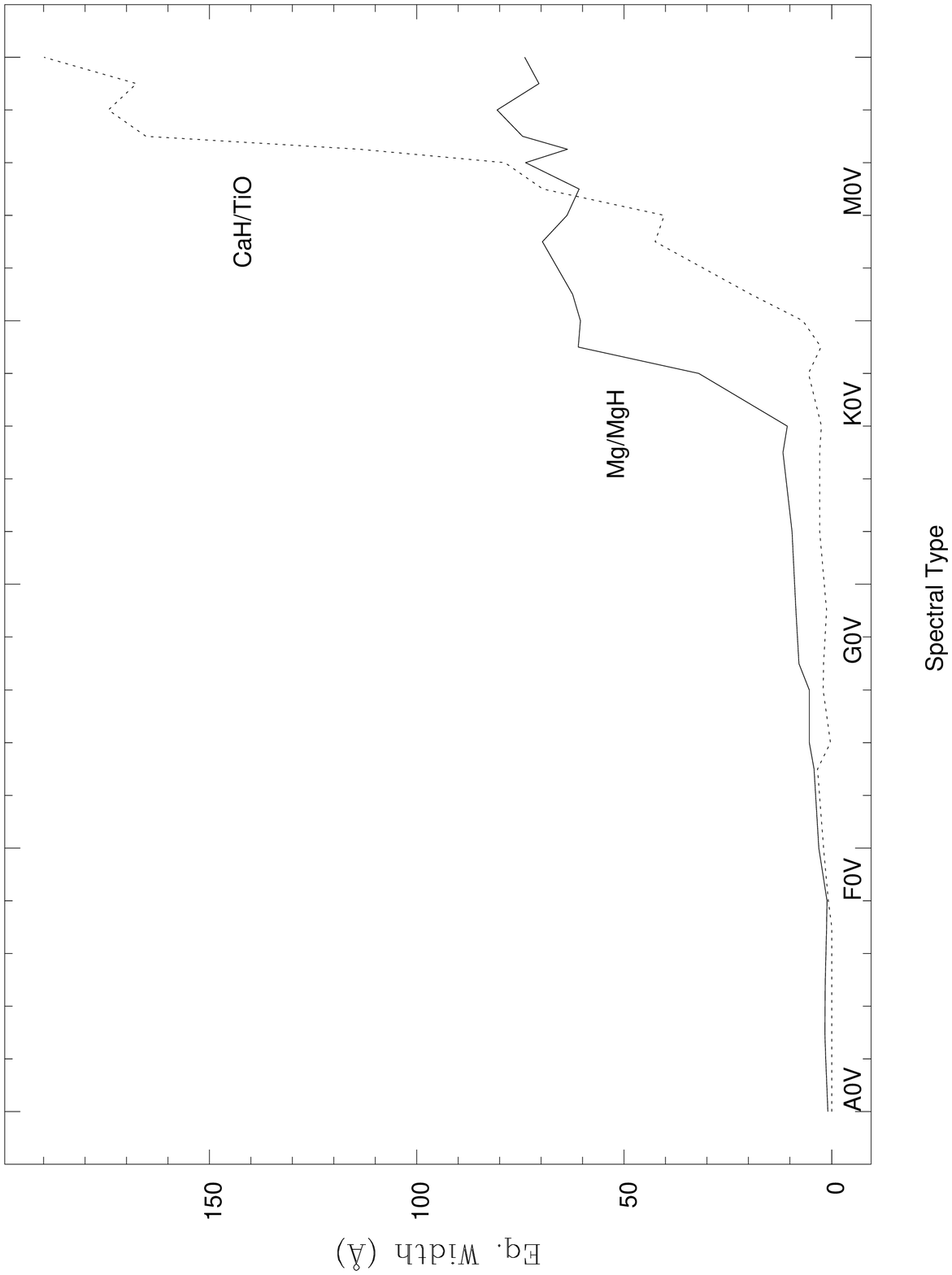]{
Equivalent width of the Mg/MgH and CaH+TiO features measured from Pickles (1998)
template spectra. The Mg/MgH feature becomes strong enough to affect the photometry
in K0 and later type
stars, whereas the CaH+TiO feature dominates at $\sim$ 6850 \AA\ for M0 and later
type stars.
\label{fig5}}

\clearpage
\begin{figure}
\plotone{kilic.fig1.ps}
\begin{flushright}
Figure 1
\end{flushright}
\end{figure}

\clearpage
\begin{figure}
\plotone{kilic.fig2.ps}
\begin{flushright}
Figure 2
\end{flushright}
\end{figure}

\clearpage
\begin{figure}
\plotone{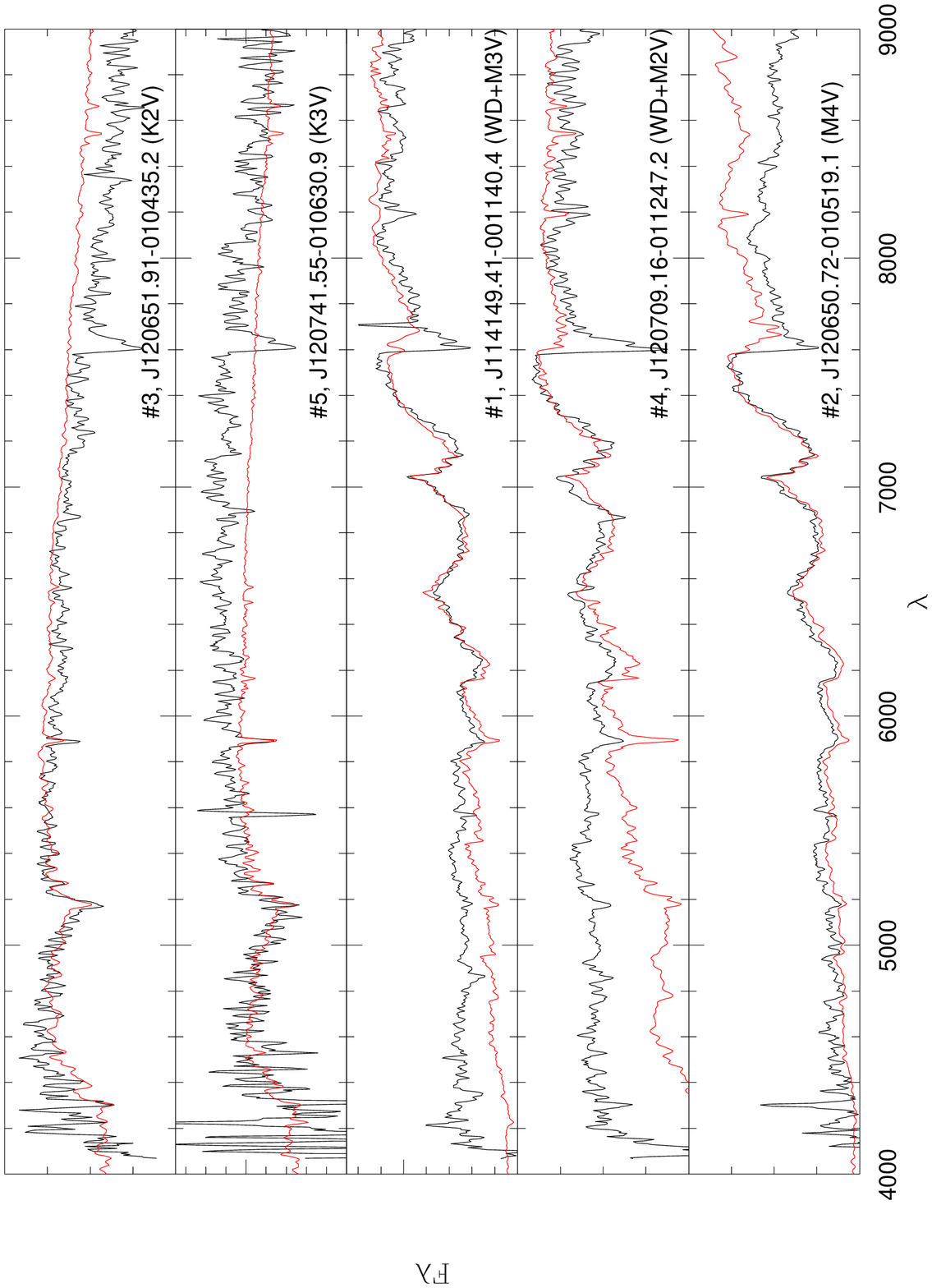}
\begin{flushright}
Figure 3a
\end{flushright}
\end{figure}

\clearpage
\begin{figure}
\plotone{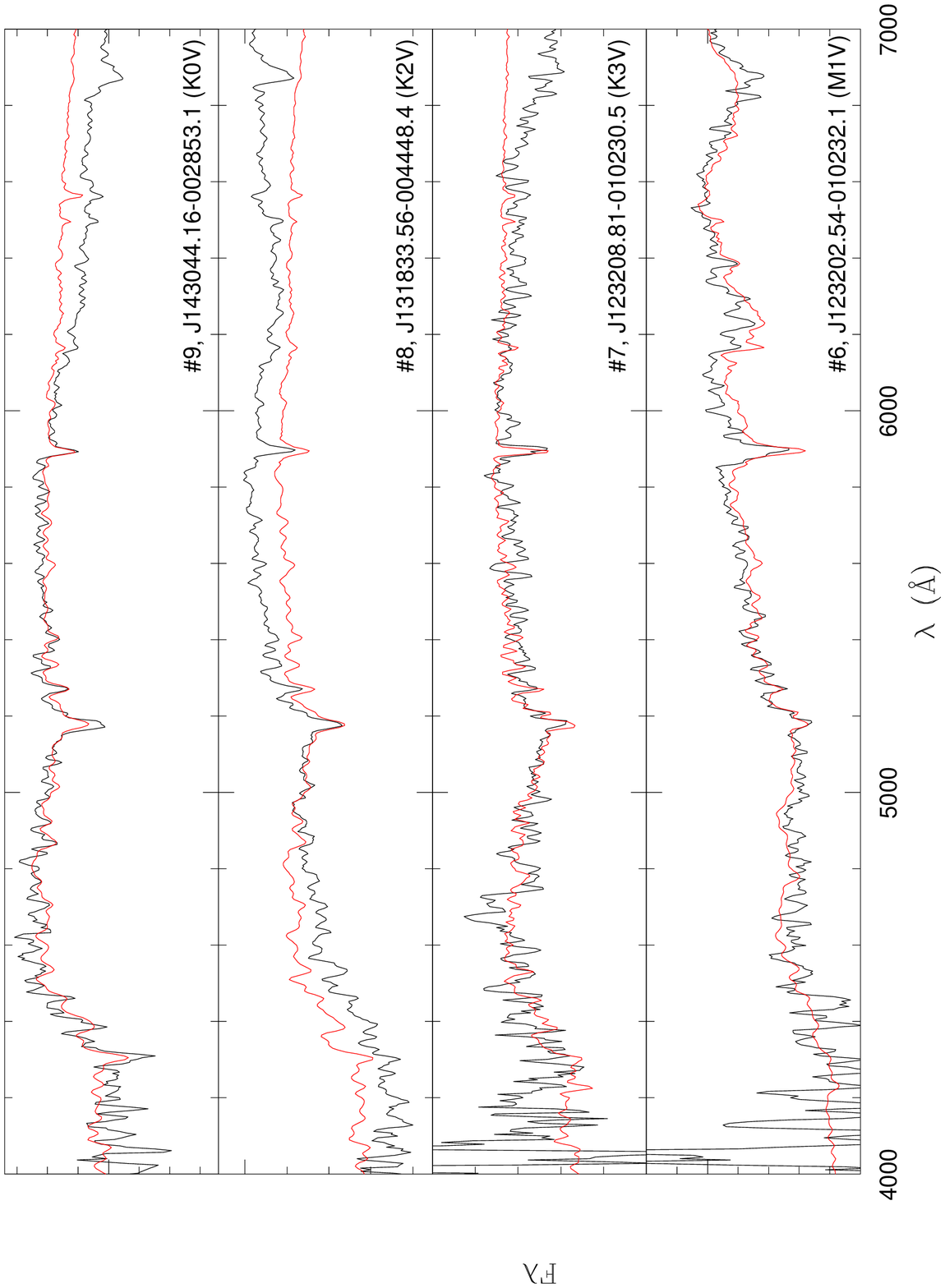}
\begin{flushright}
Figure 3b
\end{flushright}
\end{figure}

\clearpage
\begin{figure}
\plotone{kilic.fig4.ps}
\begin{flushright}
Figure 4
\end{flushright}
\end{figure}

\clearpage
\begin{figure}
\plotone{kilic.fig5.ps}
\begin{flushright}
Figure 5
\end{flushright}
\end{figure}

\end{document}